**Peak separation method for sub-lattice strain analysis at atomic resolution: Application to InAs/GaSb superlattice**

Honggyu Kim[1,2], Yifei Meng[1,2], Jean-Luc Rouviére[3] and Jian-Min Zuo[1,2*]

1. Dept of Materials Science and Engineering, University of Illinois, Urbana, Illinois 61801, USA
2. Seitz Materials Research Laboratory, University of Illinois, Urbana, Illinois 61801, USA
3. CEA/INAC/SP2M/LEMMA, 19 rue des Martyrs, 38 054 Grenoble, France

**Abstract**

We report on a direct measurement of cation and anion sub-lattice strain in an InAs/GaSb type-II strained layer superlattice (T2SLs) using atomic resolution Z-contrast imaging and advanced image processing. Atomic column positions in InAs and GaSb are determined by separating the cation and anion peak intensities. Analysis of the InAs/GaSb T2SLs reveals the compressive strain in the nominal GaSb layer and tensile strain at interfaces between constituent layers, which indicate In incorporation into the nominal GaSb layer and the formation of GaAs like interfaces, respectively. The results are compared with the model-dependent X-ray diffraction measurements in terms of interfacial chemical intermixing and strain. Together, these techniques provide a robust measurement of atomic-scale strain which is vital to determine T2SL properties.

*Corresponding Author, email: jianzuo@illinois.edu

## I. INTRODUCTION

Epitaxial heterostructures consisting of two or more semiconductors are widely used in micro- and nano-electronics such as transistors [1, 2], quantum cascade lasers [3], infrared photodetectors [4], and light emitting diodes [5]. A critical engineering parameter in designing an epitaxial heterostructure is strain, which arises from the lattice mismatch between dissimilar materials. For example, strained silicon, coupled with $Si_{1-x}Ge_x$, is used to enhance the effective carrier mobility in the channel of a metal-oxide-semiconductor field effect transistor (MOSFET). [6] The control over unconventional electronic and magnetic properties of semiconductor [7-9] and oxide heterostructures [10, 11] also relies on strain. It is experimentally challenging to measure strain in such heterostructures, especially the interfacial strain, since the region of interest is on the order of nanometer. The determination of strain in epitaxial heterostructures in general requires both high resolution and precision in order to develop a detailed understanding of strain-property relationships.

High resolution electron microscopy (HREM) has been developed as a major tool for measuring strain at high spatial resolution. Electron diffraction techniques, such as convergent beam electron diffraction (CBED)[12, 13] and nano-beam diffraction (NBD)[14, 15], are highly accurate, but their spatial resolution is limited by the electron probe size to a few nanometers. [16] Under favorable conditions, HREM images can be analyzed for constructing two dimensional (2D) strain maps using techniques such as geometrical phase analysis (GPA)[17] and peak pairs algorithm (PPA)[18]. In addition, the template matching method has been developed to measure lattice distortions for strain analysis.[19, 20] Recently, these techniques have been extended to atomic resolution Z-contrast images acquired with scanning transmission electron microscopy (STEM) using a high angle annular dark field detector (HAADF).[21, 22] Compared to HREM, the contrast in Z-contrast images is more uniform and less sensitive to thickness and defocus values, which is beneficial for quantitative strain measurement.[23]

A major challenge in using Z-contrast images for strain measurement in compound semiconductors is the intensity overlap from the neighboring atomic columns, which makes the accurate measurement of atomic column positions difficult. Even with advances in the aberration corrected STEM, which brings the image resolution to 1 Å or less, the peak intensity overlaps in the commonly observed <110> directions of the diamond or zinc blende structure (for an example see Fig. 1(a)). For the GPA analysis, Peters et al. have pointed out that because of the change in the structure factor phases in heterostructures like III-V semiconductors artifacts could arise in the strain analysis.[24] In the pattern recognition methods, e.g., template matching, the overlap of atomic column intensity gives rise to different intensity distributions and bonding distance depending on the chemical composition, and there is no single template which can be used to find the positions of every overlapped atomic columns. Here, we overcome these issues by separating the cation and anion sub-lattice using a peak-fitting based subtraction method. The strain of the separated lattices are measured using the template matching (TeMA) method.[20]

## II. EXPERIMENTAL

An InAs/GaSb T2SL grown on the GaSb substrate by molecular beam epitaxy (MBE) was studied for its lattice strain. Both InAs and GaSb belong to the zinc-blend structure. The InAs/GaSb T2SL has a period of 4.4 nm thick InAs and 2.1 nm thick GaSb of nominal composition. The InAs/GaSb based T2SLs have attracted great interest as a potential material system for the next generation infrared (IR) photodetectors operating in mid-wavelength (3~5 µm) to long-wavelength (8~12 µm) regime.[8, 25] The sample for STEM was prepared by conventional mechanical polishing using diamond lapping films, followed by Ar ion milling process at liquid nitrogen temperature to minimize the ion beam induced damages. Z-contrast images were recorded using a FEI Ultimate STEM operated at 300 kV. The microscope is equipped with a probe spherical aberration corrector and a high angle annular dark field detector (HAADF). To measure strain along the out-of-plane direction, the electron probe was scanned along the growth direction in order to reduce errors caused during the scan fly-back time.

## III. METHODS

The principle of our peak separation technique is illustrated in Fig. 1 using a dumbbell image from the GaSb substrate (cubic, a=6.0959 Å) as an example. The image was recorded along the [110] zone axis orientation. The dumbbell is formed by a pair of the cation and anion atomic columns, which are separated by a distance of 1.524 Å. The peak separation involves the following steps:

1) Each dumbbell is detected using a *dumbbell template* consisting of a pair of atomic columns. The purpose of this template is to locate every single dumbbell in the image. For this, we use the spatially averaged dumbbell over the entire image using the template matching technique.[20, 26];

2) Two Gaussian peaks are used to fit each dumbbell image using the least squares fitting method by a minimization of $\sum \left( I_i^{exp} - I_i^{fit} \right)^2$, where the sum is over all image pixels of the dumbbell. The best fit gives the fitted images for the anion and cation atomic column as shown in Fig. 1 (c) and (d), respectively;

3) By subtracting off one of the two fitted atomic peaks from the experimental image, separate anion and cation lattice images can be obtained as shown in Fig. 1 (e) and (f), respectively.

In the above steps, the choice of the dumbbell template used to locate the dumbbell position does not affect the final result of the Gaussian peak fitting, since in addition to the Gaussian peak height and width, the peak positions are also adjusted for the best fit.

In the example shown in Fig. 2 (a) from a region near the InAs and GaSb interface, two sub-lattice images are obtained for the anions and the cations, displayed in Fig. 2 (b) and (c), respectively. Figure 2 (d) is the difference intensity between the experimental image and fitted images. Only random noises are observed, demonstrating the robustness of our fitting results.

To measure strain from the sub-lattices, the position of individual atomic columns is determined using an *atomic peak template* consisting of single atomic column. Figure 2(e) shows an atomic peak template of 36 x 36 pixels obtained by spatial averaging. The sub-lattice image contains both low and high

Z atomic columns, thus both contribute to the spatially averaged peak template. We also experimented with templates based on the atomic columns of low and high Z elements, separately. The results for strain measurements are the same, which indicates that the peak shapes are approximately the same for both atomic columns. Using the atomic peak template in Fig. 2 (e), we then located the positions of every atomic columns in the InAs/GaSb T2SL using TeMA.[20]. A reference lattice was also defined for the GaSb buffer layer using the method described in ref.[19], this lattice is later used to calibrate the strain in the superlattice. Strain maps for the anion and cation sub-lattices were obtained using the e-matrix method.[20]

Next, we examined the robustness of peak separation method with template matching for atomic column determination. It is known that the template matching based peak finding using the cross-correlation function effectively filters out the noise in the Z-contrast image, creating correlation peaks which is less sensitive to the local intensity variations and noises.[20] Because of this advantage, the template matching has been widely used for pattern recognition in (S)TEM image analysis.[20, 27, 28, 29] However, it is less clear how well template matching works with the separated sub-lattice images. The GaSb substrate, where the strain or distortion of lattice is minimal, was used to examine the measurement precision, which can be defined as the standard deviation of the measured displacements. The measurement precisions based on the peak separation method along x (the scan direction) are 1.9 pm ($\sigma_x$) for the anion lattice and 2.9 pm ($\sigma_x$) for the cation lattice in Fig. 3 (a) and (b) while the measurement precisions along y (perpendicular to the scan direction) are 2.4 pm ($\sigma_y$) for the anion lattice and 2.0 pm ($\sigma_y$) for the cation lattice. Overall, the picometer precision in the measured Ga and Sb column positions demonstrates the effectiveness of our peak separation procedure and the substantially reduced overlapping peak intensities.

For comparison, we performed strain analysis using the geometrical phase analysis (GPA) algorithm.[17] This technique measures the atomic displacements in a real space image using the phase information obtained in the Fourier space, and has been previously applied to InAs/GaSb T2SLs.[22]

## IV. STRAIN IN THE InAs/GaSb T2SL AND DISCUSSIONS

Figure 4 (a) shows an atomic resolution Z-contrast image of InAs/GaSb T2SL, imaged along the [$\bar{1}$10] zone axis. The dumbbell-like features are seen for InAs and GaSb constituent layers, and because of the relatively large difference in atomic numbers between In (Z =49) and As (Z=33), or Ga (Z=31) and Sb (Z=51), the dumbbell images are asymmetric in opposite directions, which makes it easy to distinguish InAs from GaSb and vice versa.

Applying the method described above to Fig. 4 (a), we obtained separate images of cation and anion sub-lattices as displayed in Fig. 5 (a) and (b), respectively. As shown in the magnified image in Fig. 5(b), the intensity of each atomic column is well separated from the nearest neighboring atomic column. Figure 5 (c) and (d) display the out-of-lattice ($\varepsilon_{xx}$) strain maps from each sub-lattice using TeMA[20].

The strain ($\varepsilon_m$) in our strain map is measured based on the out-of-plane lattice mismatch of the film with respect to the reference, GaSb buffer layer, given by Eq. (1)

$$\varepsilon_m = \frac{a_f^\perp - a_{GaSb}}{a_{GaSb}} \qquad (1)$$

where $a_f^\perp$ and $a_{GaSb}$ (6.0959 Å) are the out-of-plane lattice constant of the deposited film and the bulk lattice constant of GaSb, respectively. Since the T2SL is lattice-matched along in-plane direction with GaSb substrate, $a_f^\perp$ can be calculated by [30],

$$a_f^\perp = a_f^{bulk}(1 - \frac{2\upsilon_f}{1-\upsilon_f}\varepsilon_f^\parallel) \qquad (2)$$

where $\varepsilon_f^\parallel = (a_f^\parallel - a_f^{bulk})/a_f^{bulk}$ is the in-plane strain of the film and $\nu_f$, $a_f^{bulk}$ are the Poisson ratio and the bulk lattice constant of the film.

Both of the cation and anion sub-lattice strain maps (Fig. 5 (c) and (d)) show distinct layer-by-layer superlattice features with compressive strained GaSb and tensile strained InAs. While the stoichiometric GaSb constituent layer grown on the GaSb substrate show zero strain, the nominal GaSb in InAs/GaSb SL is under compressive strain with the maximum strain ($\varepsilon_m$) at 0.025. This result is attributed to In segregation into the nominal GaSb during MBE growth at elevated temperature, which is consistent with the atom probe tomography result and X-ray modelling result in ref[26, 31]. Another representative feature is the tensile strain near the interface region, as indicated by red arrows in Fig. 5 (e) and (f). The intensity profile across InAs and GaSb layers in Fig. 4 (b) shows the lower intensities near the interface (marked by the red dotted line), indicating that a significant amount of GaAs bonds exist and lead to the tensile strain. The strain value alone cannot determine the exact composition of the interface bonds, thus this is the subject of future studies.

Next, we compared our method with the geometrical phase analysis (GPA) algorithm. Figure 6 (a) shows the strain map obtained from the GPA[32] of the as-recorded Z-contrast image in Fig. 4 (a) using the Bragg reflections of (002) and (220) as displayed in Fig. 6 (b). The map shows large positive $\varepsilon_m$, indicating the compressive strain at interfaces between InAs and GaSb. The compressive strain can only arises from InSb bonds, however it contradicts with the lower intensities or dark contrast near the interfaces as discussed above. The same GPA on the separated sub-lattice images yield the strain maps shown in Figure 6 (c) and (d) for the cation and anion sub-lattice, respectively. These strain maps are similar to Fig. 5(c) and (d) obtained by TeMA, albeit at a lower spatial resolution. The (002) reflection used for the GPA analysis is composition sensitive and its phase changes sign from InAs to GaSb. It has been proposed that by properly choosing Bragg spots the artifacts induced by the phase shift from the base image can be corrected.[24] Our results here show that these artifacts can be entirely avoided by using the peak separation method.

A major concern in the TEM based strain analysis is that the stresses can be significantly relaxed near surfaces of the prepared TEM sample.[33] Lattice plane bending near the specimen surfaces can cause significant change in the measured atomic peak positions and yield surface strain rather than the representative bulk strain. In a superlattice of period $\Lambda$, the strain can be analyzed based on the spatial frequency $m\Lambda$ and its Fourier components. Treacy and Gibson showed that the superlattice tends to exhibit bulk behavior at the interface, with the relaxation of each Fourier component being confined to a shallow depth $\sim\Lambda/m$ from the surfaces, even in thin specimens.[33] Away from the interface, where the low frequency Fourier components dominate, the relaxation penetrates deeper and will behave as a relaxed thin film if $t \leq \Lambda$ where t is the sample thickness. To examine the thin-film relaxation effect, we have separately determined the superlattice strain by X-ray diffraction[31].

Figure 7 plots the strain profiles obtained using TeMA and the digital model based fittings of X-ray diffraction. The strain profiles determined by X-ray diffraction are based on the free strain model (FSM) and composition-correlated strain model (CCSM), respectively. In FSM, the displacements of all 7 GaSb monolayers and 3 InAs monolayers near each interface are adjusted during fitting, while in CCSM the distance between two neighboring atomic layers are considered as an alloy structure and calculated using Vegard's law. [31] The X-ray fitting results shown in Fig. 7 were obtained using a smoothing filter, which averages the stain value at each point with two adjacent data points with a weight of 0.5. Both the STEM and X-ray diffraction results show several important strain features. First, the X-ray refinements give the average lattice constants of $d_{GaSb}$=6.190 Å and $d_{InAs}$=6.029 Å and $d_{GaSb}$=6.197 Å and $d_{InAs}$=6.023 Å for FSM and CCSM respectively, while the strain values obtained with STEM images show the average lattice constants of $d_{GaSb}$=6.185 Å and $d_{InAs}$=6.042 Å. The bulk lattice constant of InAs at 6.0583 Å is smaller than that of GaSb at 6.0959 Å, thus InAs experiences a small tensile strain when it is grown epitaxially on the GaSb substrate and the out-of-plane lattice constant of the strained InAs is expected to be 6.0174 Å. The measured $d_{InAs}$ by atomic resolution imaging and X-ray diffraction are slightly larger than the expected value for pure InAs, indicating an expansion of the nominal InAs. This result can be interpreted as the evidence of the Sb segregation into InAs during the growth, which was supported by the atom probe tomography result as well as X-ray diffraction.[31] Secondly, $d_{GaSb}$ is much larger than the lattice constant of pure GaSb, which indicates a compressive strain in the nominal GaSb layer. This can be attributed to In segregation into GaSb due to the large lattice constant of InSb (6.4794 Å). The amount of In segregation into GaSb is much larger than that of Sb segregation into InAs[31], thus it is reasonable to assume an In background in the GaSb layer. The presence of In in the nominal GaSb is also confirmed by our previous APT analysis.[31] Another notable feature is the tensile strain near the interfacial regions, which arises from the GaAs-like bonds. The tensile interfacial strain is seen by FSM and CCSM at the GaSb-on-InAs interface and the InAs-on-GaSb interface, respectively, while both are observed by atomic resolution imaging. It is known that asymmetric interface compositional grading exists in InAs/GaSb T2SLs and the InAs-on-GaSb interfaces are typically more extended than that of the GaSb-on-InAs interfaces due to Sb segregation.[26, 34] Therefore, the missing tensile strain at one of the interface by X-ray diffraction shows the limitation of the model based approach to X-ray diffraction. One possible explanation is that the exponential decay model employed in FSM is too simplified for the description of chemical bonds with high degree of chemical intermixing. Details of the FSM studies are presented in the ref[31]. All in all, both STEM and X-ray diffraction results are consistent in terms of strain values in the constituent layer and their interfaces, demonstrating the strain measured using our method is the bulk strain and less affected by the surface effects.

## V. CONCLUSION

In conclusion, we have developed a quantitative image analysis method to perform strain mapping using atomic resolution Z-contrast images. It has been found that the intensity overlap and non-uniform intensity variations in a dumbbell in Z-contrast images of III-V semiconductor heterostructures result in artifacts in strain map. To overcome these issues, we performed peak separation between sub-lattices from the experimental Z-contrast image. By applying this approach to InAs/GaSb T2SL, we demonstrated that our approach provides a basis for robust strain measurement in atomic resolution Z-contrast images, with either a real space approach (TeMA) or a Fourier space method (GPA). In particular, this method allows us to locate atomic column positions at a few picometer (< 3 pm) precision with one frame of Z-contrast image.

The analysis we carried out does not require the knowledge of chemical composition of the atomic columns. The image contrast of each atomic column helps in determining the atomic peak position, but its amplitude is not important. We note that further improvements of measurement precision should be possible with post-image processing such as averaging a series of registered images.[35, 36] The application of the image analysis tool employed here can be extended to a variety of epitaxial heterostructures in general.

**Acknowledgement**

This work was supported by U.S. Army Research Office (Grant No. Army W911NF-10-1-0524 and monitored by Dr. William Clark) through the MURI program. The STEM experiment at Grenoble was supported by Nanoscience Foundation, France. We also thank Dr. Amy Liu from IQE for providing InAs/GaSb T2SL sample.

**Figure captions**

Fig. 1. (a) GaSb dumbbell image from a Z-contrast image and (b) intensity line profiles across GaSb dumbbell. (c) anion and (d) cation fitted images obtained with Gaussian peak fitting to (a). Separate (e) anion and (f) cation lattice images obtained by subtracting the fitted images from the original image in (a).

Fig. 2. (a) Z-contrast image near interface between InAs and GaSb. (b) Anion and (c) cation lattice images. The white arrows in (a)-(c) indicate the location of interface. (d) The intensity difference between the experimental image and Gaussian fitted images. (e) Atomic peak template image used to locate the atomic column in the lattice images.

Fig. 3. The histograms of the measured displacements from the anion ((a) and (c)) and cation ((b) and (d)) sub-lattices of GaSb buffer layer for peak separation method.

Fig. 4. (a) Atomic resolution Z-contrast image of InAs/GaSb T2SL. (b) The intensity profile from the region marked by the red dotted box in (a).

Fig. 5. (a) Caion and (b) Anion sub-lattice images obtained by peak separation method. (c) and (d) Strain maps for the out-of-lattice component from (a) and (b) obtained by TeMA. (e) and (f) Strain profiles are plotted by averaging strain values of each monolayer in (c) and (d). The red arrows in (e) and (f) indicate tensile strain peaks near interfaces between InAs and GaSb.

Fig. 6. (a) Strain map obtained using GPA and Bragg spots highlighted with yellow circles in (b) FFT. (c) and (d) Strain maps from cation and anion sub-lattices using the same parameters as (a).

Fig. 7 Strain profiles obtained using STEM and X-diffraction studies. The combination of peak separation method and template matching (TeMA) for STEM images are used while the free strain model (FSM) and composition-correlated strain model (CCSM) for X-ray diffraction fitting are used to extract strain information.

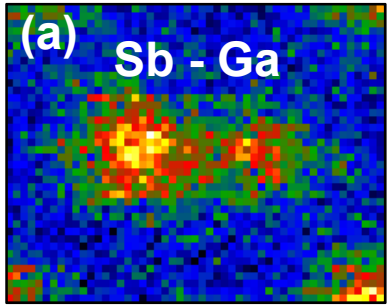
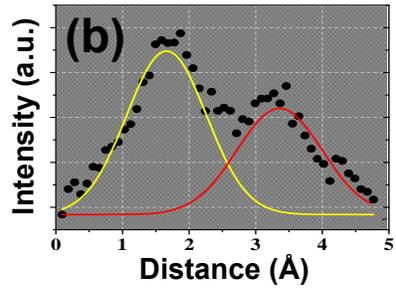
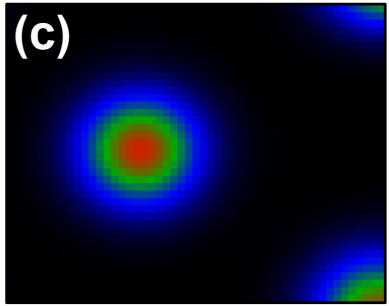
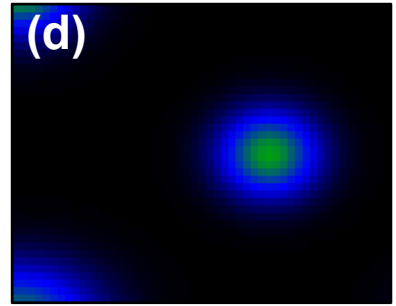
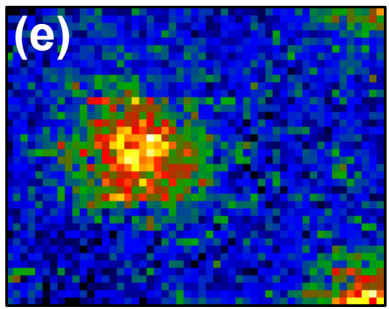
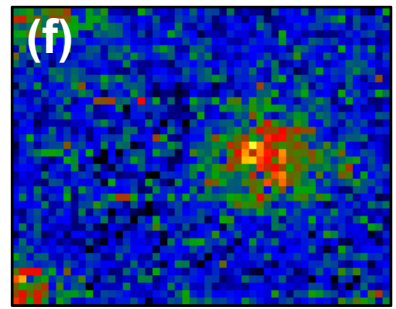

**Figure 1**

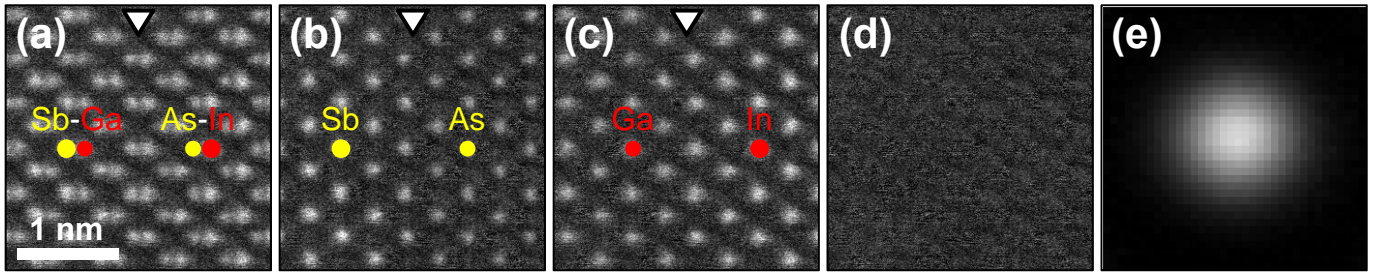

**Figure 2**

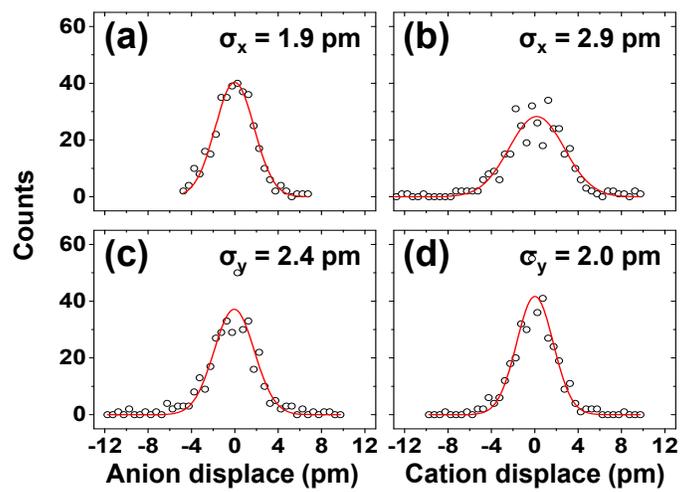

**Figure 3**

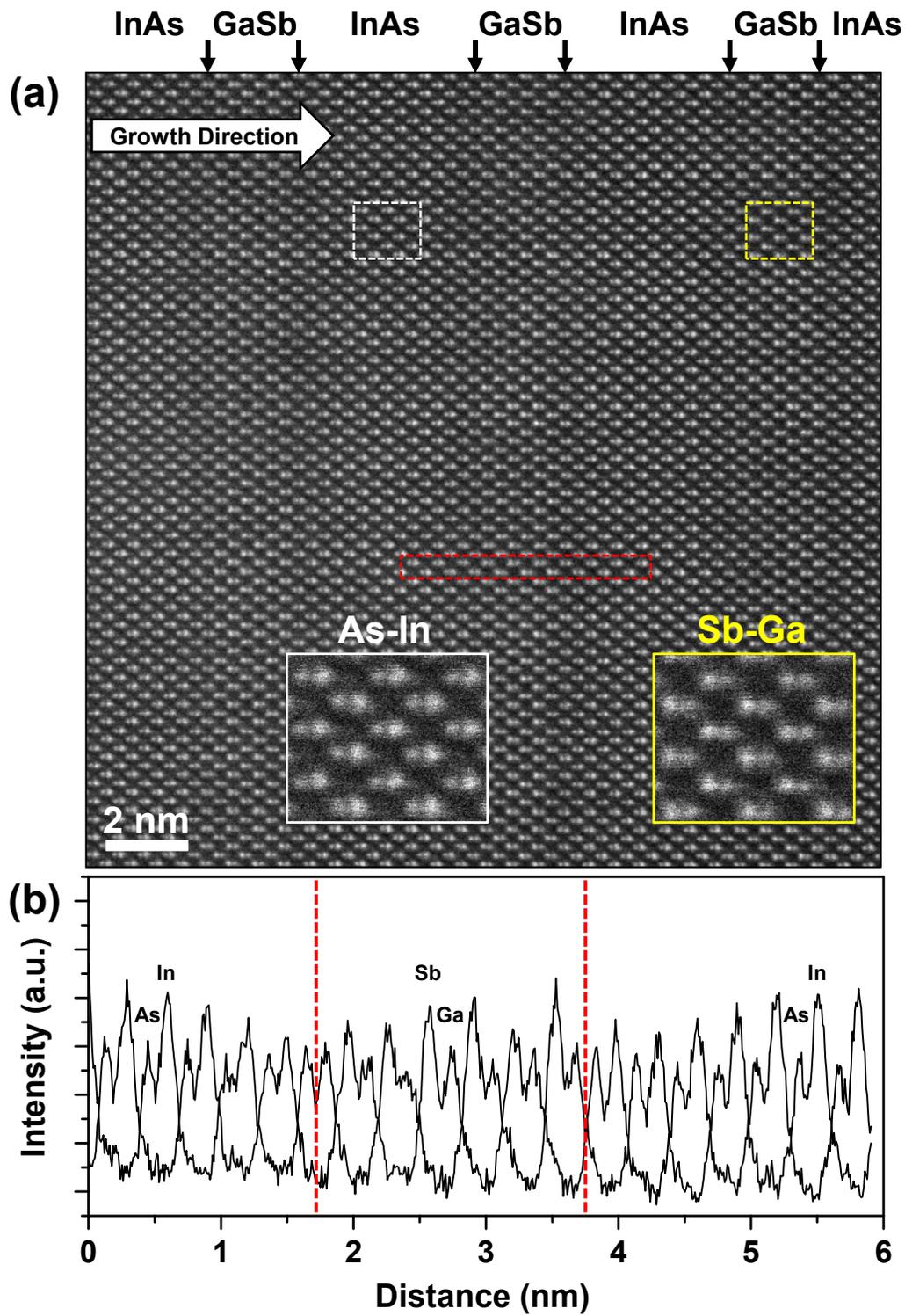

**Figure 4**

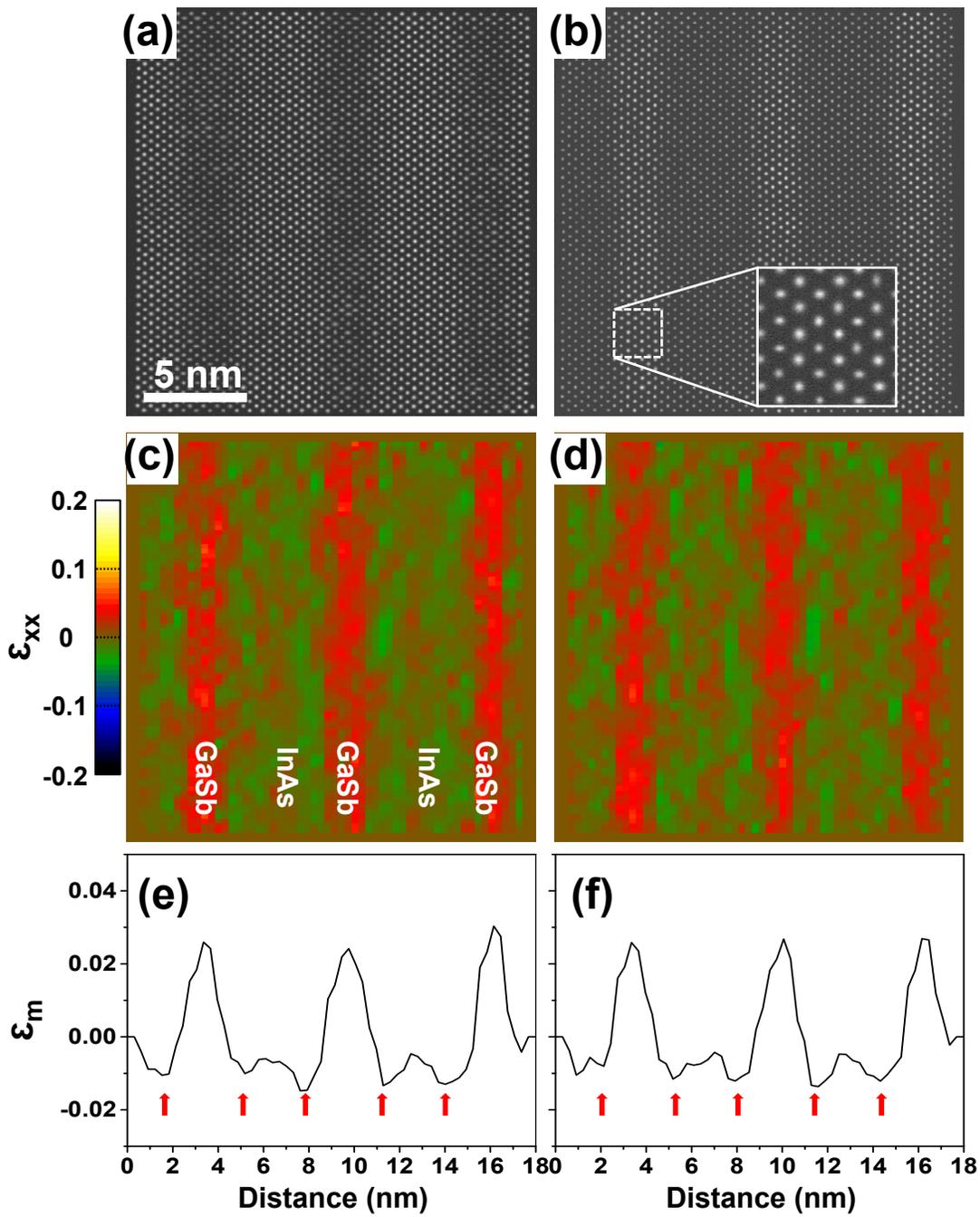

**Figure 5**

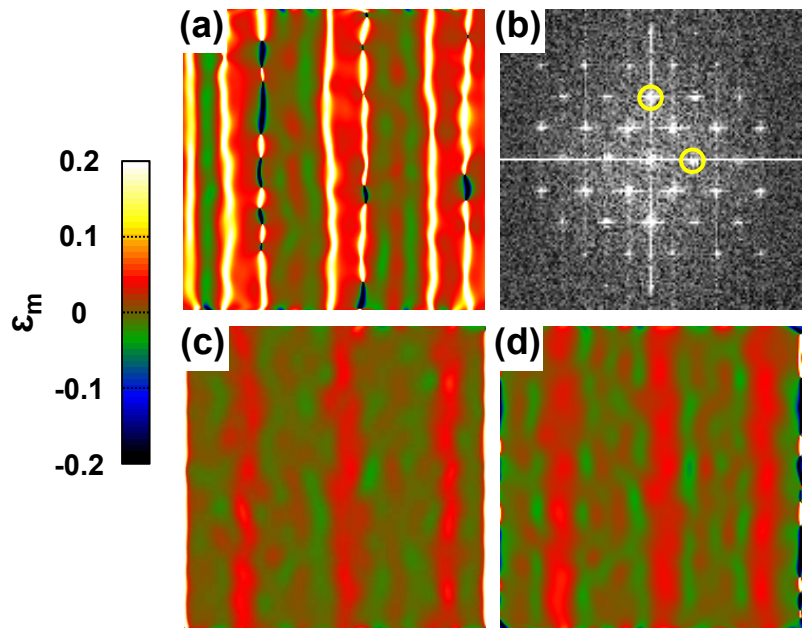

**Figure 6**

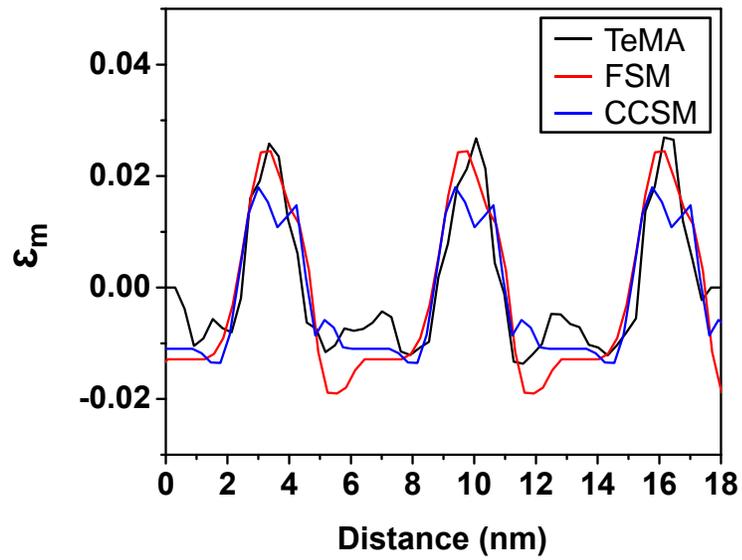

**Figure 7**